# $\sigma_h$-Broken Induced Topological quasi-BIC

Yongqi Chen[1], Chaofeng Xie[1], Tongtong Zhu[1], Weiqiang Ding[2,*] and Yurui Fang[1,*]

[1]*Center for Photophysics and Nanoscience, School of Physics, Dalian University of Technology, Dalian 116024, China.*

[2]*School of Physics, Harbin Institute of Technology, Harbin, 150001, China*

[*]*Corresponding authors: wqding@hit.edu.cn (W.D.), yrfang@dlut.edu.cn (Y.F.)*

**Abstract**

Transitions from bound states in the continuum (BICs) to quasi-BICs ($q$BICs) are typically realized by introducing in-plane asymmetry, including permittivity asymmetry ($\varepsilon$-$q$BICs) and geometry asymmetry (g-$q$BICs). Here, we demonstrate that when the in-plane symmetry is rigorously kept, the transition can also be occurred, provided the out-of-plane asymmetry is designed, which is called $\sigma_h$-$q$BICs in this work. When the $\sigma_h$ symmetry is gradually broken, the system undergoes a topological phase transition characterized by a Zak phase inversion, leading to a band inversion between quadrupole and dipole modes. This process not only enables controlled radiation coupling of BICs but also introduces a defect-immune $q$BIC regime. Our findings establish a general mechanism for engineering high-$Q$ resonances and topologically robust plasmonic cavities.

As a non-Hermitian topological non-trivial state in an open system, bound states in the continuum (BICs) have been intensively investigated in photonics. BICs often arise from the cancellation of symmetric hybridized dipolar resonances, forming dark states that are decoupled from the radiation continuum and, in theory, exhibit infinite $Q$-factors [1,2]. By breaking the symmetry, the sharp resonances have been widely exploited in applications such as nanolasers [3–7], nonlinear optics [8–16], sensing [17–20], and topological photonics [21–29].

To date, most studies on BIC-to-$q$BIC transition have focused on *breaking in-plane symmetries*, such as mirror or rotational symmetry, where lateral structural perturbations lift symmetry protection and open radiative channels [30–33]. More specifically, there two ways to introduce the asymmetry, i.e., the permittivity asymmetry ($\varepsilon$-$q$BICs) [34–39], and geometry asymmetry (g-$q$BICs) [41–50]. In these studies, the Bloch waves are only considered in the 2D plane, and the structures are considered having $\sigma_h$ symmetry even if they are on substrates [50,51]. These approaches are grounded in well-established group-theoretical selection rules, compatible with planar nanofabrication, and broadly effective in controlling radiation losses and tailoring Fano resonances. This body of work has profoundly shaped our understanding of symmetry-protected photonic modes and remains a cornerstone of high-Q design.

On the other hand, the radiation features of the $q$BICs from in-plane symmetry broken mentioned above are not flexible to tune, such as the charity and radiation direction, etc. For this reason, out-of-plane asymmetry is introduced together with the in-plane asymmetry to tune the far field light characters. For example, a slant resonant metasurface was designed to generate intrinsic chiral BIC with near-unity circular dichroism of 0.93 and high-quality factor exceeding 2,663 for visible frequencies [45]. In Ref. [24,52] , the photonic crystal slabs that radiate only towards one side of the slab is achieved with the assistance of a tilted sidewall. However, it is emphasized that the *out-of-plane asymmetry* does not determine the occurrence of the transition from BIC to $q$BIC in these studies, but contribute to the tuning of field features only [29,53].

Actually, the role of horizontal mirror symmetry $\sigma_h$, a fundamental symmetry element in many layered and metasurface and photonic crystal slabs, has not been systematically explored in this context. A natural question raises: can $q$BICs emerge from symmetry protected ideal BIC by out-of-plane symmetry broken ($\sigma_h$-broken) only with the in-plane symmetry being kept rigorously? Our answer is yes. In this Letter, we consider a plasmonic lattices that inherently support symmetry-protected BICs, as shown in FIG. 1. By systematically introducing vertical asymmetry, we find that $\sigma_h$-broken activates z-oriented radiative moments that are otherwise symmetry-forbidden, thereby opening radiative channels in the out-of-plane direction and transforming ideal BICs into $q$BICs. Full-wave simulations further verify that the $Q$-factor decreases continuously with increasing vertical asymmetry, without requiring any in-plane perturbation. These results establish $\sigma_h$ symmetry as an independent and effective degree of freedom for manipulating radiative losses in BIC-based plasmonic systems, offering a new route toward symmetry-controlled resonance engineering in non-planar metastructures.

As shown in FIG. 1, we begin by illustrating our general strategy through the evolution from an array of metallic cylinders to an array of metallic cones, forming a photonic crystal slab composed of a square lattice with period $a$ in both the $x$ and $y$ directions, which supports Bloch-type solutions of the form [54,55] $A(r)=A_{kn}(r)=u_{kn}(\rho)\exp[i(\mathbf{k}_\parallel \cdot \rho)]\exp[i(k_z z)]$, $u_{kn}(\rho+a_i)=u_{kn}(\rho)$, where $\boldsymbol{\rho}=(x,y)$ represents the in-plane coordinate. $\mathbf{k}_\parallel=(k_x,k_y)$ denotes the in-plane Bloch wavevector, and $k_z$ is the out-of-plane wavevector component determined by the boundary conditions. The continuity of the tangential components of the electric and magnetic fields at the upper and lower interfaces gives $E_{1y}\cos(k_z z)=E_{2y}\exp(-\kappa z), k_z E_{1y}\sin(k_z z)=\kappa E_{2y}\exp(-\kappa z)$, where $\kappa=\sqrt{k_\parallel^2-\varepsilon_2 k_0^2}$ is the transverse decay constant of the evanescent field in the surrounding dielectric $\varepsilon_2$, and $k_0 = \omega/c$ is the free-space wavenumber. From which non-trivial dispersion relations for both guided and radiative modes can be obtained, with $\kappa \cos(k_z d) \mp k_z \sin(k_z d) = 0$ for TE, $\sigma_z = \pm 1$ for TM.

Even modes of the cylindrical unit cell remain symmetry-protected and do not couple to free-space radiation (FIG.1(a)), giving rise to bound states in the continuum (BICs) (FIG 1(d)). In the limit of large thickness, the structure supports standing waves along the z-direction, while in thinner slabs, the two interfaces hybridize, leading to strong coupling between the two sides. As the pole radius of the upper ends reduce, the even modes becomes radiative when the radius smaller then the cutoff size of the mode (FIG. 1(b)). In this condition, the TE and TM modes of the photonic crystal couple togather [56]. Continue to reduce the pole upper radius will make the unit cell a cone (FIG 1(c)), in which case a Zak protecting phase will appear (FIG 1(e)).

Now we investigate the evolution of the photonic crystal from BIC to $q$BIC. Consider a metallic cylinder unit cell, whose modes can be expressed as

$$E_z(\rho, \varphi) = A_m I_m(k_0 \kappa_m \rho) \cos(m\varphi), \tag{1}$$

$I_m$ is the modified Bessel function of the first kind, and $\kappa_m$ is the radial propagation constant determined by the boundary condition at the pillar surface. The dispersion relation of the cylinder [57] is shown in FIG. 2(a). In this geometry, the $m = 2$ mode corresponds to a symmetry-protected BIC. As the pillar radius decreases, higher-order modes $m \geq 1$ gradually reach their cutoff frequencies and become radiative. Thus when the cylindrical lattcie is reshaped into a truncated cone with upper and lower radii $r_u < r_d$, higher-order modes confined in the lower side become radiative once the local radius falls below the cutoff threshold. Further reducing $r_u$ continuously transforms the structure into a cone, whose eigenmodes are described by [58]

$$\Phi = \sum_{k,m} A_{k,m} \cos(m\varphi) \frac{1}{r^{1/2}} \cos(h\eta \ln \frac{r}{r_0} + \delta) P_{1/2+ik}(\cos\theta) \qquad (2)$$

where $r$ and $\theta$ denote the radial and polar coordinates of the conical geometry, $h$ is the longitudinal wave parameter, and $P^m_{-1/2+ik}(\cos\theta)$ is the associated Legendre function of complex order $-1/2+ik$, representing the angular field dependence on the cone surface. The dispertion relation shows very similar behavior as cylinder (FIG.2(b)). The radius $r$ corresponding to position $z$ reduces as $z$ increasing, which causes increasing wavevector (untial infinity principally). The fundamental mode $m = 0$ exhibits pronounced field localization near the cone apex, while higher-order modes evolve into radiative channels as they approach the tip. As a consequence, the BIC originating from the $m = 2$ mode (and other high order even modes) in the cylindrical geometry transitions into a quasi-BIC once the horizontal mirror symmetry $\sigma_h$ is broken and the upper radius becomes smaller than the mode cutoff size (FIG. 1(e)). In the conical limit, the structure exhibits extreme field concentration with an ultralow optical mode volume.

As the cone height continues to decrease, the resonant frequencies of the $m = 1$ (dipole) and $m = 2$ (quadrupole) modes, shift in opposite directions—the dipole mode red-shifts while the quadrupole mode blue-shifts—leading to a band inversion and the emergence of a topologically nontrivial photonic state. Meanwhile, the $m = 0$ mode, which cannot be excited in TE condition in the cylindrical unit cell photonic slab, corresponds to a vertically polarized dipole, giving rise to a highly confined localized resonance within the photonic crystal slab.

To verify the above theoretical picture, we experimentally and numerically investigate a plasmonic photonic crystal that follows the same geometric evolution from $\sigma_h$-symmetric to $\sigma_h$-broken configurations. FIG. 1(d) illustrates the $\sigma_h$ symmetry photonic crystal lattice consisting of a square lattice of cylindrical poles (lattice constant $a = 400$ nm, radius $r = 0.37a$) perforated in an aluminum film. This structure preserves horizontal mirror symmetry ($\sigma_h$) and therefore should support symmetry-protected BICs. The cylindrical lattice gradually reshapes into the topological lattice FIG. 1(e), completing the geometric evolution, in which the $\sigma_h$ symmetry is broken while the periodicity of the lattice remains unchanged. This process effectively transforms the photonic crystal slab from a symmetric to an asymmetric configuration, allowing weak coupling between the originally protected BICs and free-space radiation. The geometric evolution of a single unit cell during the schematical transformation process is following FIG. 1(a-c).

FIG. 3(a) highlights the geometric and band evolution from symmetry-protected BICs to q-BICs induced by mirror-symmetry breaking and structural compression. Initially, the cylindrical lattice exhibits $D_{4h}$ symmetry. As the upper surface becomes gradually compressed, and sharp nanocone-like tips emerge. At a critical deformation of $r_u = 0.08a$, the system undergoes a topological transition, entering a $C_{4v}$-symmetric configuration with a nontrivial topology.

For the cylindrical lattice, four bands appear within the spectral window of interest. At the $\Gamma$ point, the lowest-frequency band ($\Gamma_1$) corresponds to a quadrupolar mode ($m = 2$), as indicated by the blue dashed, giving rise to a symmetry-protected BIC with an ideally infinite Q-factor (FIG.3(a), top). The middle two bands form a doubly degenerate pair ($\Gamma_{2\,3}$) protected by $D_{4h}$ symmetry. When the structure approaches the critical configuration, the breaking of $\sigma_h$ symmetry causes the TM and TE modes to no longer be strictly eigenfrequencies, and the modes begin to mix, leading to the mixing of a quartet accidental degenerate mode near the $\Gamma$ point and the closing of the bandgap. (FIG.3(a), mid.)

As the nanocone compression exceeds the critical threshold (low radius $r_d = 0.37a$, top radius $r_u < 0.08a$, (FIG.3(a), bottom)), the degeneracy is lifted and a topological band gap opens, signaling a band inversion between the quadrupole-like and dipole-like modes. This transition is accompanied by a significant redistribution of the near-field energy: the quadruple field becomes concentrated near the cone apex and gradually leaks into the far field. As shown in FIG. 3(b), the electric-field distributions of the upper (high-frequency) and lower (low-frequency) bands reveal a clear inversion of $E_z$ parity when the conical geometry reaches the compression limit, confirming the occurrence of band inversion and the emergence of a topological plasmonic phase (see detail in Supplementary Materials, Note 1).

The resulting Q-factor evolution is shown in FIG. 3(c). In the cylindrical state, the $\sigma_h$-protected BIC exhibits a theoretically divergent Q at the $\Gamma$ point, whereas in the critical and topological regimes, the breaking of $\sigma_h$ symmetry leads to finite radiation losses, forming quasi-BICs with $Q \geq 10^4$. As shown in Fig. 3(d), the quality factor Q reduces with the cone angle $\theta$ as the geometry is continuously deformed, accompanied by the breaking of $\sigma_h$ symmetry. Once the cone angle exceeds a critical value $tan(\theta) = 0.768$, a pronounced decrease in Q is observed, indicating that the conical geometry induces strong field concentration near the apex and enhanced radiation into the far field.

The topological band flip, corresponding to a shift in the geometrical phase of the hybrid band. This $\sigma_h$ symmetry-breaking topological transition is captured by the 2D Zak phase, for which the general expression of the x and y components is given by [59,60]:

$$\theta_i(k_j) = -\mathrm{Im}\left[\ln\left(\prod_{n=1}^{N}\langle u(k_{n,i}, k_j)|u(k_{n+1,i}, k_j)\rangle\right)\right], \quad (i \neq j;\ i,j \in \{x,y\}) \tag{3}$$

$|u_n(k)\rangle$ denotes the Bloch eigenfunction on the $n^{th}$ band with wave vector $k$. Selecting the $\Gamma$ point of the topological lattice as the inversion center, as the truncated cone continuously evolves into a cone, with the top radius $r_u$ decreasing from *0.13a* to *0*, a pronounced Zak phase transition is observed, with $\Gamma_1$ and $\Gamma_4$ having (0, 0) and, ($\pi$, $\pi$) confirming the band inversion (FIG.3(a), bottom). This topological protection ensures that the resulting quasi-BIC remains robust against structural imperfections and fabrication-induced disorder (see Supplementary Materials, Note 2).

To further verify the theoretical mechanism of BIC evolution, we simulate the angle-resolved reflectivity spectrum of cylindrical lattices evolving to topological lattices with $\sigma_h$ symmetry breaking.

For the cylindrical lattice, the reflection spectrum exhibits a vanishing dip at the $\Gamma$ point of the lower band, as shown in FIG 4(a), confirming the existence of a symmetry-protected BIC. In the truncated cone lattice, the lower-band BIC remains visible at the $\Gamma$ point (FIG 4(b)). Importantly, the introduction of $\sigma_h$-symmetry breaking gives rise to additional spectral features: a distinct surface state emerges in the middle band near the $\Gamma$ point, providing clear evidence of starting coupling to m = 1 mode.

Upon entering the critical regime, the spectrum FIG. 4(c) reflects the fourfold accidental degeneracy predicted in the band analysis, corresponding to an ill-defined Zak phase. Further compression ($r_u = 0$) lifts this degeneracy and reopens the bandgap. In this regime, a dark BIC mode emerges in the upper band ($\Gamma_4$) around FIG. 4(d), accompanied by a Fano-type asymmetric line shape in the reflection spectrum. This behavior directly demonstrates the transition of the symmetry-protected BIC into a finite-Q dipolar BIC along the out-of-plane direction.

These spectral results provide strong numerical evidence that the observed *q*BICs originate from the controlled breaking of $\sigma_h$ symmetry, and their radiative leakage and line-shape evolution are consistent with the theoretical band analysis in FIG. 3.

In conclusion, breaking the $\sigma_h$-symmetry through controlled out-of-plane geometric perturbations drives the transition from symmetry-protected BICs to quasi-BICs in photonic crystals, providing a unified framework for understanding and engineering high-Q resonances. The induced in-plane Bloch wave mismatch coupled to the vertical mode achieves longitudinal radiation leakage while maintaining strong in-plane confinement. This geometry-driven approach reveals the fundamental mechanism of BIC generation and breaking in metallic systems and establishes symmetry control as a universal design principle. By shifting the focus from particle-level radiation directionality to symmetry-guided mode engineering, this strategy enables precise manipulation of *q*BICs and topological transitions, offering a robust pathway toward complex and tunable high-Q photonic and plasmonic responses.

## Acknowledgements

The authors thank the supporting of the National Natural Science Foundation of China (Grant No. 12274054, 12574414).

## Author contribution

Y.F. conceived the idea and directed the project. W.D. directed the physics picture. Y.C. performed the main experiment and the main FEM calculations. C.X. performed the dispersion relation calculation. T.Z provided inspired discussion. All of the authors discussed the results and basic clue for writing. Y.C. wrote the manuscript. All authors revised the paper.

## Conflicts of interest

The authors declare no competing financial interest.

**Figures and Captions**

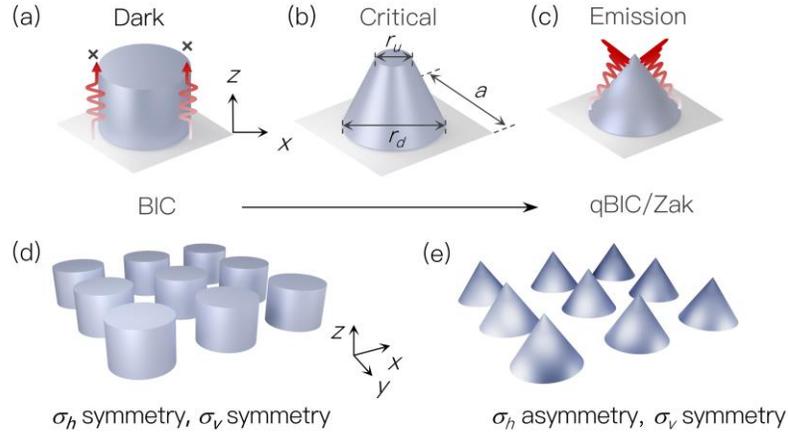

**FIG. 1.** Evolution from cylindrical to conical unit cells and the corresponding mode characteristics in a photonic crystal slab. (a) Illustration of patterns in a metal cylinder array. Regarding the even mode (m = 2) under C2 symmetry within the plane, it is decoupled from the plane wave solution, forming BICs. Standing wave radiation in the z-direction is prohibited and confined to the near field, exhibiting symmetric protection along the z-axis. (b) Schematic of the truncated-cone array. The broken mirror symmetry between the top and bottom surfaces enables weak coupling of the confined mode to radiation, resulting in q-BICs. (c) Schematic of the conical array. Further breaking of the $\sigma_h$ symmetry causes strong field accumulation near the cone apex, where the confined energy eventually couples into free space. (d) Schematic of the cylindrical photonic crystal lattice consisting of a square lattice of cylinders (lattice constant $a$ = 400 nm, radius $r$ = 0.37$a$) in an aluminum slab. The structure preserves mirror symmetry with respect to the horizontal plane ($\sigma_h$) and thus supports symmetry-protected bound states in the continuum (BICs). (e) By geometric evolution, the cylindrical lattice is reshaped into a topological lattice with broken $\sigma_h$ symmetry.

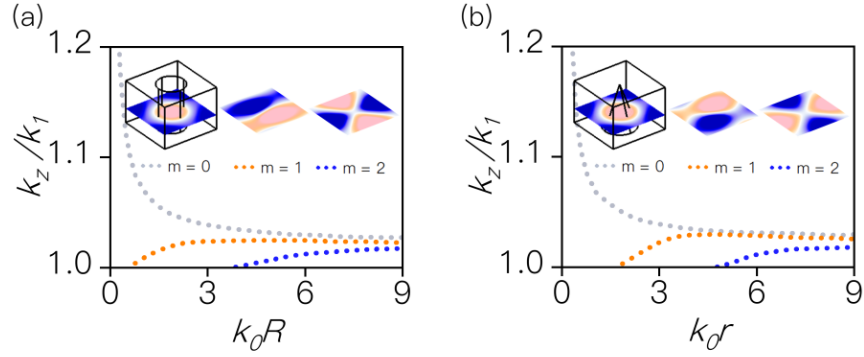

**FIG. 2. Cylinder and cone lattice dispersion curves** (a) Dispersion relations for the modes supported by a periodic array of metallic cylinders, showing the fundamental ($m = 0$), dipole ($m = 1$), and quadrupole ($m = 2$) branches. The inset illustrates the electric-field distributions of the three modes within unit cell. R represents the cylinder radius, $k_1$ is the wavevector in surrounding medium. (b) Dispersion relations for the conical array, where higher-order modes appear at larger in-plane wavevectors compared with the cylindrical case, indicating their gradual transition from localized guided modes to radiative continua as the cone geometry opens. r represents the radius of the cone crossing section.

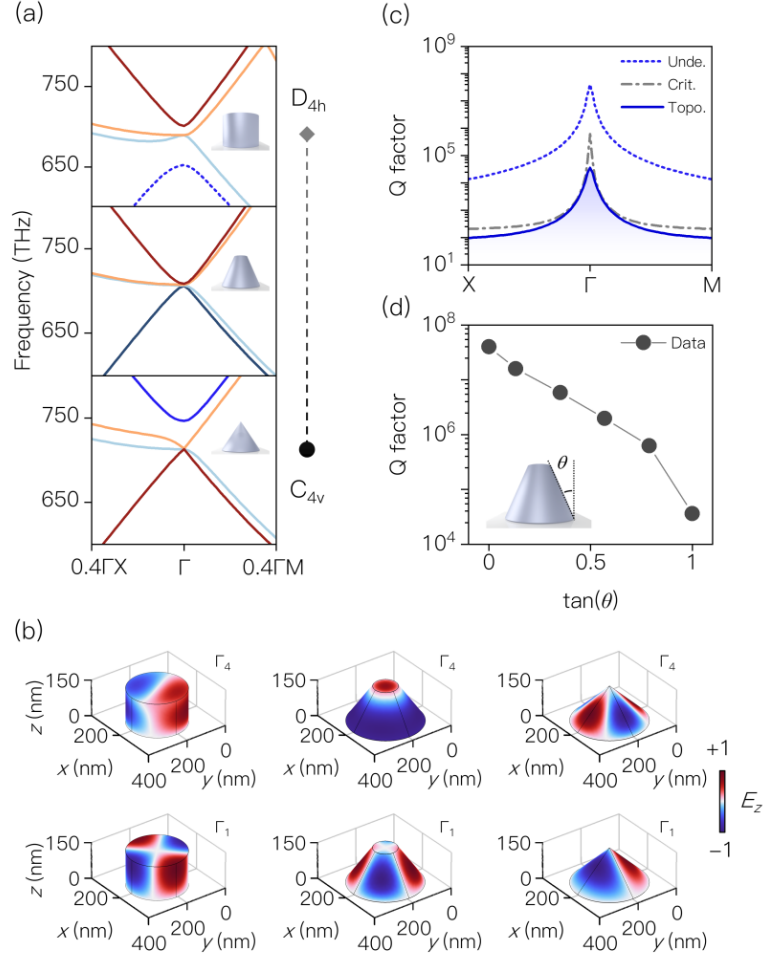

**FIG. 3. Band structure evolution under $\sigma_h$ symmetry breaking.** (a) Band structures of the lattices calculated under periodic boundary conditions. (Top) The cylindrical lattice supports a symmetry-protected BIC at the Γ point, where the lowest band (Blue dashed line) corresponds to the quadruple mode (m = 2), and the doubly degenerate $\Gamma_{2\,3}$ bands are protected by $\sigma_h$ symmetry. Upon symmetry breaking and compression, mode hybridization among $\Gamma_1$, $\Gamma_4$, and $\Gamma_{2\,3}$ closes the band gap near the critical configuration and leads to a topological band inversion (mid. and bot.). (b) Electric-field $E_z$ distributions at the Γ point for the upper (high-frequency) and lower (low-frequency) bands. As the conical geometry is continuously compressed from a cylindrical shape to its limiting configuration, a clear inversion of the even parity of $E_z$ is observed, evidencing band inversion and the emergence of a topological plasmonic phase. (c) Evolution of the Q-factor corresponding to the quadruple mode. The cylindrical lattice exhibits an ideal BIC with divergent Q at the Γ point, while the critical and topological lattices show finite-Q quasi-BIC and multi-BIC states ($\sim 10^4$) due to controlled radiation leakage and enhanced far-field coupling. (d)

Dependence of the Q factor on the cone angle $\theta$ as the geometry evolves from a cylindrical to a conical configuration.

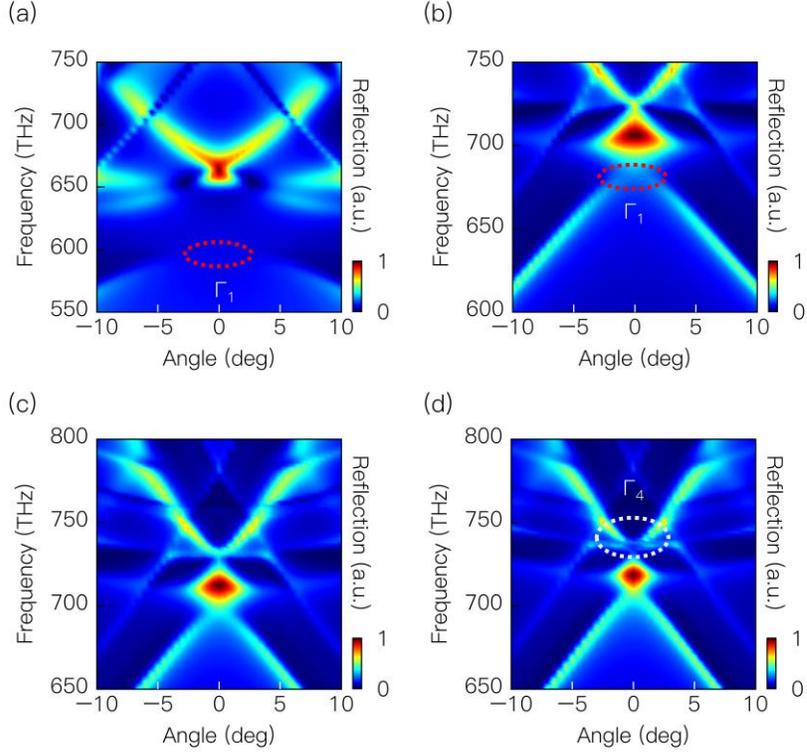

**FIG. 4. Angle-resolved spectra showing BIC-to-*q*BIC transition.** (a) Cylindrical lattice ($D_{4h}$): vanishing reflection at the Γ point confirms the existence of a symmetry-protected BIC in the lower band (red dashed circle). (b) Truncated cone lattice ($C_{4v}$): The BIC persists at the low frequency band, while mirror symmetry breaking introduces an additional surface state near the middle band. (c) Critical lattice: fourfold accidental degeneracy occurs at the Γ point, corresponding to an ill-defined Zak phase. (d) topological lattice: the degeneracy is lifted and a dark dipolar BIC emerges in the upper band (746 THz, blue dashed circle), accompanied by a Fano-type asymmetric resonance line shape.

# Supplementary Materials for

# $\sigma_h$-Broken Induced Topological quasi-BIC

Yongqi Chen[1], Chaofeng Xie[1], Tongtong Zhu[1], Weiqiang Ding[2,*] and Yurui Fang[1,*]

[1]*Center for Photophysics and Nanoscience, School of Physics, Dalian University of Technology, Dalian 116024, China.*

[2]*School of Physics, Harbin Institute of Technology, Harbin, 150001, China*

*Corresponding authors: wqding@hit.edu.cn (W.D.),   yrfang@dlut.edu.cn (Y.F.)*

**Supplementary Note 1: Evolution of Near-Field Distributions and Parity Inversion**

To further elucidate the physical mechanism behind the topological transition, we examine the evolution of the electric field ($E_z$) distributions as the geometric symmetry is continuously broken. As the nanocylinder is deformed into a nanocone, the vertical and radial symmetries are modified, leading to a significant redistribution of the electromagnetic near-field.

In the initial cylindrical configuration (Fig. S1, left column), the modes exhibit characteristic multipolar symmetries. Specifically, the dipole-like and quadrupole-like modes are well-defined, with $E_z$ parity reflecting the high symmetry of the structure. However, as the nanocone compression increases—modeled by decreasing the top radius $r_u$ while maintaining a fixed bottom radius $r_d = 0.37a$—the mode degeneracy is gradually lifted. When the top radius reaches the critical threshold ($r_u < 0.08a$), a topological band gap opens. This geometric limit forces the near-field energy to concentrate significantly at the cone apex. As shown in the rightmost column of Fig. S1, the quadrupolar field becomes tightly confined at the tip before eventually leaking into the far field as the compression reaches its limit.

The most striking evidence of the topological phase transition is the inversion of $E_z$ parity. By comparing the upper (high-frequency) and lower (low-frequency) bands in Fig. S1, it is evident that the spatial distributions of the dipole-like and quadrupole-like modes are interchanged upon passing the critical geometric threshold. This field redistribution confirms a band inversion, which serves as a definitive signature of the emergence of a topological plasmonic phase in the nanocone system.

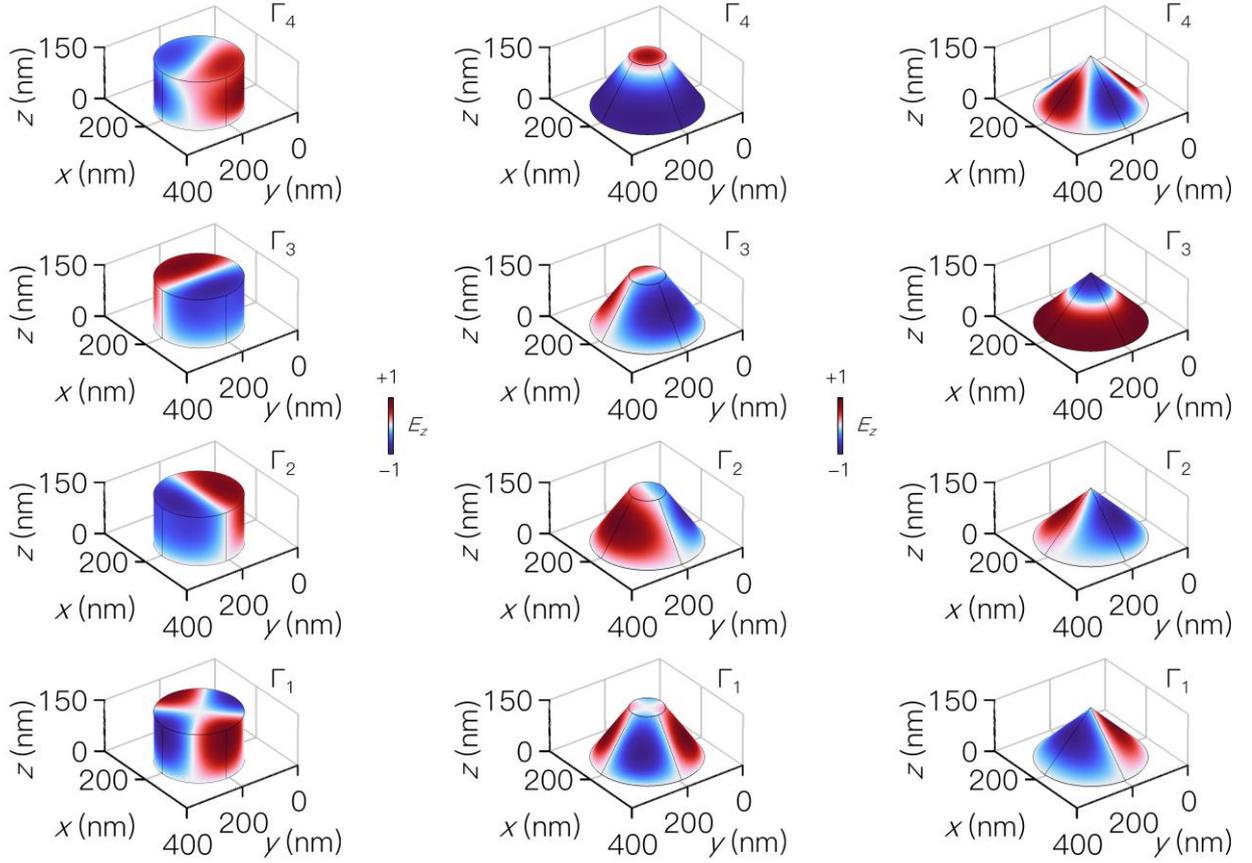

**Figure S1 Evolution of modal $E_z$ profiles during nanocone compression.** This figure visualizes the $E_z$ near-field evolution from a nanocylinder to a sharp nanocone as the top radius $r_u$ decreases below the $0.08a$ threshold. A clear parity inversion between the upper ($\Gamma_4$) and lower ($\Gamma_1$) bands is observed at the compression limit, with field energy concentrating at the cone apex. This spatial redistribution provides direct visual confirmation of the topological band inversion and the emergence of the topological plasmonic phase discussed in the main text.

## Supplementary Note 2: Robustness of Topological quasi-BICs against Disorder

To further validate the topological nature of the observed resonance, we investigate the robustness of the quasi-bound states in the continuum (quasi-BICs) within a finite 6×6 nanocone supercell. As established in the main text, the transition from truncated cones to sharp cones ($r_u \to 0$) induces a Zak phase transition and band inversion at the $\Gamma$ point. This topological transition provides a unique mechanism to protect the symmetry-protected BICs against external perturbations.

In Fig. S2a, we present the near-field $E_z$ distribution of the quasi-BIC mode for a pristine 6×6

lattice. The mode exhibits a highly collective and symmetric field profile, characteristic of the Γ-point eigenfunction. The longitudinal electric field is strongly localized at the apex of each nanocone, maintaining the phase coherence across the entire supercell.

To test the topological protection, we introduce structural disorder by randomly displacing the nanocone positions and varying their heights and radii (Fig. S2b). In conventional non-topological photonic systems, such high-$Q$ resonances are typically extremely sensitive to fabrication-induced disorder, often resulting in significant mode scattering or frequency shifts.

As shown in Fig. S2b, despite the presence of significant structural impurities and lattice irregularities, the $E_z$ field distribution remains largely intact. The quasi-BIC mode persists with minimal distortion to its spatial envelope and phase relationship. This persistence is a direct consequence of the topological band inversion described by the Zak phase. Because the mode is protected by the global topological properties of the lattice (band inversion between $\Gamma_1$ and $\Gamma_4$), it remains robust against local perturbations that do not close the topological band gap. This robustness confirms that our nanocone platform is highly suitable for practical applications in nanophotonics and sensing, where structural imperfections are unavoidable.

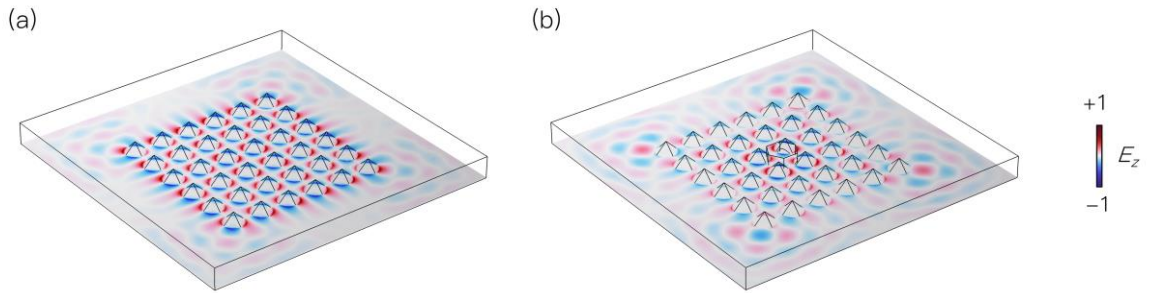

**Figure S2 Topological robustness of the quasi-BIC mode.** (a) Simulated $E_z$ field distribution for a pristine 6×6 nanocone supercell, exhibiting the regular periodic profile of a symmetry-protected BIC at the Γ point. (b) $E_z$ distribution in a disordered supercell featuring random structural perturbations; the stability of the mode profile demonstrates the topological protection afforded by band inversion. Red and blue regions represent normalized positive and negative $E_z$ phases, respectively, highlighting the robust localization of the quasi-BIC against fabrication-induced disorder.